\documentstyle[floats,aps]{revtex}
\begin{document}
\input epsf

\font\tenmib=cmmib10
\font\eightmib=cmmib10 scaled 800
\font\sixmib=cmmib10 scaled 667
\newfam\mibfam
\textfont\mibfam=\tenmib
\scriptfont\mibfam=\eightmib

\draft

\twocolumn[\hsize\textwidth\columnwidth\hsize\csname  
@twocolumnfalse\endcsname
\title{Phase separation and enhanced charge-spin coupling
near magnetic transitions}

\author{ Francisco Guinea$^1$, Guillermo
G\'omez-Santos$^2$
and Daniel P. Arovas$^3$}
\address{
$^1$Instituto de Ciencia de Materiales, CSIC,
Cantoblanco, 28049 Madrid, Spain \\
$^2$ Departamento de F{\'\i}sica de la Materia Condensada
and Instituto Nicol\'as Cabrera, 
Universidad Aut\'onoma de Madrid. Cantoblanco. E-28049 Madrid , Spain\\
$^3$Department of Physics, University of California at San Diego,
 La Jolla CA 92093}

\date{\today}

\maketitle

\begin{abstract}
The generic changes of the electronic compressibility in systems
which show magnetic instabilities is studied.
It is shown that, when going into the ordered phase, the compressibility
is reduced by an amount comparable to the its original value,
making charge instabilities also possible. We discuss, within this
framework, the tendency towards phase separation of the double exchange
systems, the pyrochlores, and other magnetic materials.
\end{abstract}

\vskip2pc]

\narrowtext
\section{Introduction}
The theoretical possibility of phase separation (PS) in magnetic systems was
first discussed in connexion to the Hubbard model for itinerant
ferromagnetism and antiferromagnetism\cite{V74}. Recent 
arguments suggest that PS is also likely in other 
magnetic materials, such as those in which magnetism is due
to double exchange 
interactions\cite{N96,RHD97,AG98,Yetal98,Detal98,AGG98}.
Spin polarons, which can be viewed as a
manifestation of PS on a small scale, have been
analyzed in relation to the pyrochlores\cite{ML98}.
Finally, there is an extensive literature on PS
in two dimensional doped antiferromagnets
(see, for instance\cite{D96,HM97,Cetal98,IFT98,CBS98,Cetal98b}), 
although there is no
definitive consensus on its existence.

A variety of different experiments show features
which are compatible with electronic phase segregation, at least
on small scales, near magnetic transitions. 
For instance, near an antiferromagnetic transition there is
ample evidence for the formation of charged stripes,
both in the cuprates\cite{Tetal95} and in the nickelates\cite{Cetal93}.
In the ferromagnetic 
manganites, various experiments suggest the existence of
polarons near the Curie temperature\cite{Tetal97}, which, as
mentioned above, can be viewed as phase separation on
small scales. In addition, and also in the 
ferromagnetic manganites (i. e. La$_{1-x}$Ca$_x$MnO$_3$,
$x \approx 1/3$),
inhomogeneous textures\cite{Uetal99} and hysteretic effects\cite{Betal99}
have been reported near $T_c$. Note that we are discussing
the relatively simple case of the optimally
doped, {\it ferromagnetic} compounds,
avoiding the complications which arise between the 
competition between antiferromagnetism, charge
ordering and double exchange in these
materials. Finally, there is evidence for
inhomogeneous textures in
ferromagnetic pyrochlores\cite{Aetal99}.
The purpose of this paper is to show the theoretical 
foundations which make likely the existence of electronic 
phase segregation near magnetic transitions. We do not pretend to
give an exhaustive list of experimental evidence which
support this view.

In the following, we analyze how a magnetic transition 
influences the electronic compressibility. A general framework
is presented in the next section. Simple applications to
variations of the Hubbard model are presented in section III.
Then, it is shown that the same framework predicts the existence
of PS in the pyrochlores (section IV). Section V
includes in the same framework the known results about  PS
in double exchange models, in the light of the present framework.
Section VI analyzes the role of long range interactions, and some
physical quantities likely to be affected by a reduced compressibility.
Finally, the main conclusions of our work are presented in section VII.

\section{General features of the electronic compressibility
near a magnetic phase transition.}
The formation of an ordered phase induces a decrease in the 
free energy of the material, usually called the condensation energy.
Because of it, quantities which depend on the variation
of the free energy with temperature, such as the specific heat,
show anomalous, non analytic behavior at the critical temperature.
For instance, the specific heat shows an abrupt reduction at
$T_c$ in mean field theories. 
The phase transition can also be tuned by varying $n$, the electronic
content (per unit cell), in many systems leading to phase
diagrams like that shown in fig. (\ref{figpiro2},\ref{defig1},\ref{defig2}). 
Thus, one expects an anomalous dependence of the free energy on
electronic density as the phase boundary is crossed by changing
the electronic concentration. 

We can get a simple estimate of the effect by using a standard
Ginzburg-Landau expansion for the free energy:
\begin{eqnarray}
{\cal F} ( s ) &= &\frac{c}{2} \int ( \nabla s )^2 d^D r
+ \frac{a [ T - T_c ( n ) ]}{2} \int s^2 d^D r
+ \nonumber \\ &+ &\frac{b}{4} \int  s^4 d^D r + {\cal F}_n ( n )
\label{GL}
\end{eqnarray}
where $s$ is the electronic magnetization and $n$ is the electronic
density. We now neglect spatial fluctuations, and obtain a mean field
approximation to ${\cal F}$:
\begin{equation}
{\cal F}_{MF} = \frac{a [ T - T_c ( n ) ]}{2} s^2 +
\frac{b}{4} s^4 + \frac{( n - n_0 )^2}{2 \tilde{\kappa}_0}
\label{MF}
\end{equation}
where we have expanded the dependence of the free energy of the
paramagnetic phase on $n$. $\tilde{\kappa}_0$ is the (scaled) electronic
compressibility in the paramagnetic phase, and $n_0$ defines the equilibrium
density in the absence of magnetism. 
When $ T < T_c$, the magnetization is:
$s = \frac{a [ T_c ( n ) - T ]}{b}$ and the free energy
becomes:
\begin{equation}
{\cal F}_{MF} = - \frac{a^2 [ T_c ( n ) - T ]^2}{4 b} +
\frac{( n - n_0 )^2}{2 \tilde{\kappa}_0}
\label{freeen}
\end{equation}
Let us now fix the temperature $T$ and expand this expression 
around the density $n_c$ such that $T = T_c ( n_c )$.
We obtain:
\begin{equation}
{\cal F}_{MF} \approx - \frac{a^2}{4 b} \left( \frac{\partial
T_c}{\partial n} \right)^2 
( n - n_c )^2 + \frac{( n - n_0 )^2}{2 \tilde{\kappa}_0}
\end{equation}
And by taking derivatives, we have:
\begin{equation}
\tilde{\kappa}^{-1} \equiv
\frac{\partial^2 {\cal F}_{MF}}{\partial n^2} =
\left\{ \begin{array}{lr} \frac{1}{\tilde{\kappa}_0} &T_c < T \\
- \frac{a^2}{2 b} \left( \frac{\partial T_c}{\partial n}
\right)^2 + \frac{1}{\tilde{\kappa}_0} &T < T_c \end{array} \right.
\label{compress}
\end{equation} 
The compressibility has a jump at the transition. The origin of this 
discontinuity is the same as that in the specific heat. Moreover,
it is reasonable to expect that this anomaly will be enhanced 
when fluctuations in the critical region are taken into account.

Alternatively, and keeping $T$ fixed, one can minimize 
eq.(\ref{MF}) first with respect to $n$.  By expanding
$T_c ( n )$ around $n_0$, we find:
\begin{equation}
n \approx  \left\{ \begin{array}{lr} n_0 &T_c < T \\
n_0 + \frac{a \tilde{\kappa}_0 s^2}{2} \frac{\partial T_c}{\partial n}
&T < T_c \end{array} \right.
\end{equation}
and, inserting the value of $n$ in the free energy when
$T < T_c$:
\begin{equation}
{\cal F} \approx \frac{a [ T_c ( n_0 ) - T ]}{2} s ^2
+ \frac{b}{4} s^4 - \frac{ a^2 \tilde{\kappa}_0}{8} \left(
\frac{\partial T_c}{\partial n} \right)^2 s^4
\end{equation}
The dependence of $T_c$ on density leads to a negative quartic
term in the dependence of the free energy on the magnetization. 
When $ b / 4 -  ( a^2 \tilde{\kappa}_0 ) ( \partial T_c /
\partial n )^2 / 8 < 0$, the magnetic transition becomes first
order. This condition is equivalent to saying that the effective
compressibility, defined in eq.(\ref{compress}) becomes negative.
Thus, PS near the transition can be thought of as arising from  
the transmutation of a continuous phase transition into a first
order one by the introduction of an additional field,  the
density $n$, which is a well known possibility
in statistical mechanics\cite{K98}.
The new feature found in a magnetic transition is that the correction
to the compressibility can easily be comparable to the initial
compressibility. The latter is determined by the density of states
at the Fermi level in the paramagnetic phase. In typical magnetic systems,
the transition is driven by a coupling constant which is  
of the order of the inverse of the density of states. Finally, the
dependence of the critical temperature on the electronic density
 depends on the change of the coupling constant with variations in
the density of states, which is of the same order (see examples below).
Thus, no fine tuning of parameters is required to obtain corrections
to the compressibility of the order of the compressibility itself.

\section{Phase separation in Hubbard like models}
\subsection{Ferromagnetic transitions}

We first consider a one band Hubbard model  with nearest neighbor
ferromagnetic exchange couplings:
\begin{equation}
{\cal H} = \sum_{k,s} \epsilon_k c^{\dag}_{k,s} c_{k,s}
+ \sum_i U n_{i,\uparrow} n_{i,\downarrow} 
\label{hubbard}
\end{equation}
where $i$ is a site index,
$n_{i,s} = \langle c^{\dag}_{i,s} c_{i,s} \rangle$, 
$\vec{s}_i = \langle \sum_{s,s'} c^{\dag}_{i,s} 
\vec{\sigma}_{s,s'} c_{i,s'} \rangle$ and $\vec{\sigma}_{s,s'}$ 
are Pauli spin matrices.
This model (with appropriate additions) has been invoked as a reasonable starting point for the
study of itinerant ferromagnetism\cite{Vetal97}.
Within a mean field approximation, the Stoner criterium gives a
feromagnetic instability for $ U  {\cal D} ( \epsilon_F ) \ge 1 $,
where $ {\cal D} ( \epsilon )$
is the density of states per spin at energy $\epsilon$. 
Let us assume that in the paramagnetic phase there are $n_0$ electrons
per site, with chemical potential $\mu_0$:
\begin{equation}
n_0 = 2 \int^{\mu_0} {\cal D} ( \epsilon ) d \epsilon
\end{equation}
If we shift the two spin bands by $\pm \delta$, the induced
polarization $s = n_{i,\uparrow} - n_{i,\downarrow}$ satisfies:
\begin{equation}
\delta = \frac{s}{2 {\cal D} ( \mu_0 )} + \frac{ {\cal D}'^2 ( \mu_0 ) s^3}{16
 {\cal D}^5 ( \mu_0 )} - \frac{ {\cal D}'' ( \mu_0 ) s^3 }{48 {\cal D}^4 ( \mu_0 )}
\end{equation}
where $ {\cal D}'$ and $ {\cal D}''$ stand for the derivatives with respect to energy
of the density of states. Because of the lack of electron-hole
symmetry when these derivatives are finite, the chemical potential
in the polarized state is shifted:
\begin{eqnarray}
\mu - \mu_0 = - \frac{ {\cal D}' ( \mu_0 ) \delta^2}{2 {\cal D} ( \mu_0 )} +    
\frac{ {\cal D}' ( \mu_0 ) {\cal D}'' ( \mu_0 ) \delta^4}{4 {\cal D}^2 ( \mu_0 )} -            
\frac{ {\cal D}'^3 ( \mu_0 ) \delta^4}{8 {\cal D}^3 ( \mu_0 )}  \nonumber \\
-\frac{ {\cal D}'''( \mu_0 ) \delta^4}{24 {\cal D} ( \mu_0 )}                     
\label{emu}                
\end{eqnarray}
The ground state energy, at zero temperature, can be written as:
\begin{eqnarray}
{\cal F}_{s} &= &E_0 + \frac{s^2}{4 {\cal D} ( \mu_0 )} +                    
\frac{ {\cal D}'^2 ( \mu_0 ) s^4}{64 {\cal D}^5 ( \mu_0 )} -                             
\frac{  {\cal D}'' ( \mu_0 ) s^4}{192 {\cal D}^4 ( \mu_0 )} \nonumber \\ &- 
&\frac{U  s^2}{4}
\label{energyhubbard}
\end{eqnarray}  
where $E_0 = 2 \int^{\mu_0} \epsilon  \;{\cal D} ( \epsilon ) d \epsilon$.
This is the Ginzburg-Landau expansion needed to study the
phase transition as function of electronic density at zero temperature.  
The system becomes ferromagnetic when $ {\cal D} ( \mu_c ) 
 U  \ge 1 $. We can expand the quadratic term in the magnetization
around $n_c = 2 \int_0^{\mu_c} {\cal D} ( \epsilon ) d \epsilon $ as:
\begin{equation}
\frac{1}{4} \left[ U  - \frac{1}{ {\cal D} ( \mu )} \right]  \approx
\frac{ {\cal D}' ( \mu_c ) ( n - n_c )}{8 {\cal D}^3 ( \mu_c )}
\end{equation}
which leads to the following scaled inverse compressibility, as the transition is
approached from the ordered side:
\begin{equation}
\tilde{\kappa}^{-1} = \tilde{\kappa}_0^{-1}  
- \frac{1}{2 {\cal D} ( \mu_c ) }
\frac{1}{1 - \frac{D'' ( \mu_c ) {\cal D} ( \mu_c )}{3 {\cal D}'^2 ( \mu_c )}}
\end{equation}
where  
$\tilde{\kappa}_0^{-1} = (2 {\cal D} ( \mu_c ))^{-1}+ (2 U_c )^{-1}  $,  and    
$U_c = ( {\cal D} ( \mu_c ) )^{-1}$ is the critical coupling for the
transition to take place.
The compresibility is negative if
$  {\cal D}'' ( \mu_c ) {\cal D} ( \mu_c ) / (3 {\cal D}'^2 ( \mu_c )) > 1 / 2 $.
 In particular, near a saddle point in the 3D
dispersion relation, we have 
$ {\cal D} ( \epsilon ) = {\cal D}_0 - c \sqrt{| \epsilon - \epsilon_{SD}|}$. This implies 
$    \lim_{\epsilon \rightarrow \epsilon_{SD}  } 
\{ {\cal D}'' (\epsilon ) {\cal D}( \epsilon ) / {\cal D}'^2(\epsilon) \} 
\rightarrow +\infty$,
and the system is always unstable versus PS.

It is interesting to note that, using a different formalism, phase separation
has been shown to appear
near ferromagnetic phases of the Hubbard model in two dimensions\cite{H99},
in good agreement with the picture presented here.

\subsection{Doped antiferromagnets}
It is well known that the Hubbard model (eq.[\ref{hubbard}]
 at half filling, in a bipartite lattice and with
nearest neighbor hoppings only, has an antiferromagnetic (AF) ground 
state, except in one dimension. The main physical features of
this state, a charge gap, long range magnetic order and low
energy spin waves, are well described using standard mean field
techniques and the Random Phase Approximation. Recent work shows
that, in the presence of a static magnetization, the opening of
a charge gap occurs while the quasiparticle residues remain 
finite, at least in the infinite dimension limit\cite{CK98},
further supporting the validity of a mean field ansatz.
Alternatively, a calculation of the ground state energy in the limit of
large dimensions and small values of
$U/t$ can be obtained by standard perturbation theory around
a mean field symmetry breaking state\cite{D91}. This approach
does suggest the existence of PS near
half filling\cite{D96}. In the following, we analyze the
stability of homogeneous mean field solutions in arbitrary
dimensions, following a different approach to that in\cite{D96}.

The mean field solution at half filling is straightforward to
extend to finite fillings. Hartree Fock calculations show that
there are self consistent 
homegeneously doped solutions for a finite range
of fillings around half filling, in two and three 
dimensions\cite{H85,P66}.  As reported earlier, these solutions
acquire a negative contribution to the electronic compressibility
near the transition boundaries, making PS possible.
In the following, we analyze the
stability of homogeneous mean field solutions in arbitrary
dimensions, following a different approach to that in\cite{D96}.

The AF distortion  shifts the mean
 field levels,
$\epsilon \rightarrow \rm{sgn}(\epsilon)\sqrt{\epsilon^2 + \Delta^2}$,
 opening a gap and
leading to the  following staggered magnetization:
\begin{equation}
\bar{s} \equiv \sum_{\sigma=\pm 1} \sigma \; c^{\dagger}_{i,\sigma}c_{i,\sigma}=
-2 \int_{-W_0}^{\tilde{\mu}} d\epsilon \; {\cal D} (\epsilon) 
{\rm{sgn}(\epsilon)\Delta \over \sqrt{\epsilon^2 + \Delta^2}}  
\end{equation}
where ${\cal D} (\epsilon) = {\cal D} (-\epsilon)$ and $\tilde{\mu}$ 
are, respectively, the density of states (per spin) and Fermi level in 
the paramagnetic phase,
with filling factor $n = 2 \int_{-W_0}^{\tilde{\mu}} d\epsilon {\cal D} 
(\epsilon)$, and half-bandwidth $W_0$.
The energy is given by
\begin{equation}
{\cal E} =
2 \int_{-W_0}^{\tilde{\mu}} d\epsilon \; {\cal D} (\epsilon) 
\; \rm{sgn}(\epsilon) \; \sqrt{\epsilon^2 + \Delta^2}
 + \Delta \bar{s} + {U\over4}(n^2- \bar{s}^2) 
\end{equation}
 with the order parameter satisfying the following self-consistency
requirement:
\begin{equation}
{1\over U}=-\int_{-W_0}^{\tilde{\mu}} d\epsilon \; {\cal D} (\epsilon) \;
{\rm{sgn}(\epsilon) \over \sqrt{\epsilon^2 + \Delta^2}}
\label{eqaf1} 
\end{equation}
Both energy and self-consistency equation can be obtained from the minimization
of the following functional of the order parameter ${\cal F}(\Delta)$, which we take as the
starting point for the stability analysis:
\begin{equation}
{\cal F}(\Delta) =   2 \int_{-W_0}^{\tilde{\mu}} d\epsilon \;{\cal D} (\epsilon) 
 \; \rm{sgn}(\epsilon)  \; \sqrt{\epsilon^2 + \Delta^2}
+ {U \over 4} n^2  + {{\Delta^2} \over U} 
\end{equation} 
Away from half-filling, a critical value $ U_c $ is required for the instability
to take place, with $ U_c(n) = -1 / \int_{-W_0}^{\tilde{\mu}}
d\epsilon \; {{\cal D} (\epsilon) \over \epsilon} $. 
Close to the transition line $U_c(n)$, a Ginsburg-Landau
analysis is straightforward, leading to:
\begin{equation}
{\cal F}(\Delta) =  
f_0 + f_2  \;\Delta^2+ f_4 \; \Delta^4 
\end{equation} 
with
\begin{eqnarray}
f_0 &=& 
2 \int_{-W_0}^{\tilde{\mu}} d\epsilon \; \epsilon
\; {\cal D} (\epsilon)+{U \over 4} \; n^2  \\
f_2 &=& {1 / U} - {1 / U_c(n)}  \\
f_4 &=& -{1 \over 4}\int_{-W_0}^{\tilde{\mu}} d\epsilon \;{{\cal D} (\epsilon) 
\over \epsilon ^3} 
\end{eqnarray} 

The scaled inverse compressibility satisfies:
\begin{equation}
\tilde{\kappa}^{-1}=
\frac{\partial^2 {\cal E}}{\partial n^2} =
\left\{ \begin{array}{lr} {1 \over 2 {\cal D}(\tilde{\mu})} + {U \over 2 } 
,&U \rightarrow
 U_{\rm c}^{-} \\
{1 \over 2 {\cal D}(\tilde{\mu})} + {U \over 2 } 
- { ( \partial_n f_2  )^2 \over 2 f_4} 
,&U \rightarrow
 U_{\rm c}^{+} \end{array} \right.
\end{equation} 

Following the transition line on the AF side to the limit of
small doping, one has ${(\partial_n f_2  )^2 \over 2 f_4} \rightarrow 
 1 / {\cal D}(\tilde{\mu}) $, leading to
$ \tilde{\kappa}^{-1} \rightarrow -{1 \over 2 {\cal D}(\tilde{\mu})} $.
Therefore we have proven 
the existence of PS in the vicinity of $U_c(n)$, at least close to half
filling.

In fact, PS is not restricted to the previous region. It is also present
around half-filling, $ n\rightarrow 1^{\pm }$, for {\em all} values of $U$,
as we now show.
Close to half-filling and at finite $U$, the compressibility satisfies
\begin{equation}
\tilde{\kappa}^{-1} = 
U /2 - \left[ 2 \Delta^2 \int_{-W_0}^{0} d\epsilon
{{\cal D}(\epsilon)  \over (\epsilon^2 + \Delta^2)^{3/2} } \right] ^{-1}
\end{equation}
and,  given the  following inequality (see eq. \ref{eqaf1}),
\begin{equation}
2 \Delta^2 \int_{-W_0}^{\tilde{\mu}} d\epsilon
{{\cal D}(\epsilon)  \over (\epsilon^2 + \Delta^2)^{3/2} }
<
2 \int_{-W_0}^{\tilde{\mu}} d\epsilon
{{\cal D}(\epsilon)  \over \sqrt{\epsilon^2 + \Delta^2} }
={2 \over U}
\end{equation}
one obtains 
\begin{equation}
 \tilde{\kappa}^{-1} < 0
\end{equation}
proving PS around half-filling for all values of $U$, as previously stated.
This analysis indicates that PS is a very robust feature of the AF
instability at the mean field level. In fact, for a flat model density of
states, numerical results suggests that the entire AF region has negative
compressibility.

The  study of this section can be
extented to the case of a diverging density of states at half-filling: 
${\cal D}(\epsilon) \sim |\epsilon|^\alpha$, $(0 \leq \alpha < 1)  $. 
The existence of PS is also proved in this case,
at least around the transition line $U_c(n\rightarrow 1)$.
In particular, this includes 
the mean field solution of the Hubbard model in the square lattice, where
the density of states has a van Hove logarithmic divergence. 
The conclusions of our analysis are consistent
with the tendency towards inhomogeneous solutions found in mean
field studies of the Hubbard model\cite{ZG89,PR89,Sc90,Vetal91}.

\section{Pyrochlores}
The pyrochlores are metallic oxides,
Mn$_2$Tl$_2$O$_7$, which show various anomalous
transport and magnetic properties, including colossal magnetoresistence.
The magnetism is mostly due to the spins of Mn ions. The electronic carriers
are assumed to come from a wide s band from Tl orbitals, with a
very low occupation. These electrons are coupled ferromagnetically
to the Mn spins\cite{Setal96,Aetal99}. We can describe the coupled system
by the hamiltonian\cite{ML98}:
\begin{eqnarray}
{\cal H} &= &\sum_{k,s} \epsilon_k c^{\dag}_{k,s} c_{k,s} - 
\nonumber \\ &- &J \sum_{ij} \vec{s}_i
\vec{s}_j - J' \sum_i c^{\dag}_{i,s} \vec{\sigma}_{s s'} c_{i,s'} \vec{s}_i
\label{hamiltonian}
\end{eqnarray}
where $J$ and $J'$ are positive (ferromagnetic). The model undergoes
a transition to an ordered where the local spins have a finite magnetization,
$s = | \langle \vec{s} \rangle | \ne 0$, and the electron gas is polarized,
$\bar{s} = | \langle c^{\dag}_{i,s} \vec{\sigma}_{s s'} c_{i,s'} \rangle |
\ne 0$. 

In principle, the entire issue of PS in the pyrochlores could be settled using
results of Majumdar-Littlewood(ML)\cite{ML98}. These authors have shown that  
magnetic polarons (i.e. self-trapped, polarized electrons in a magnetic bubble
of core spins embedded in 
a paramagnetic environment) are the stable configuration of carriers in the
paramagnetic phase at low densities. Now we will show that the existence of
polarons in the paramagnetic  phase is a sufficient condition for the
thermodynamic 
instability of the
uniform phase of the standard mean field analysis.

\begin{figure} [!t]
\centering
\leavevmode
\epsfxsize=8cm
\epsfysize=8cm
\epsfbox {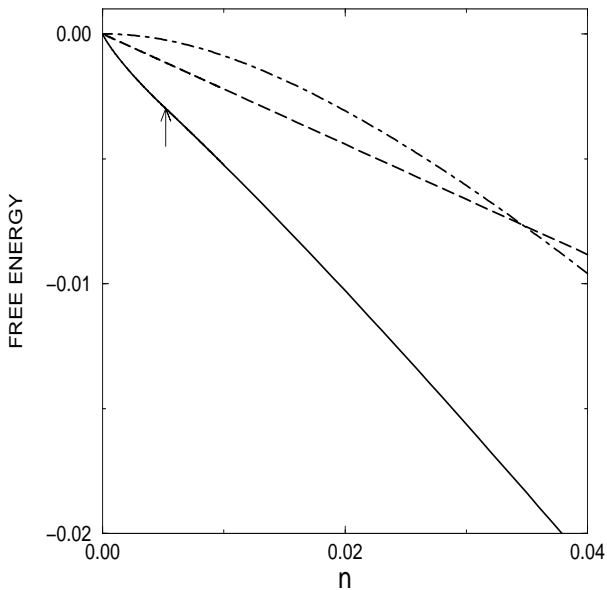}
\caption[]
{\label{figpiro1}
Free energy versus electron concentration for 
the exact mean field treatment (continuous line), the polaronic ansatz (dashed
 straight line), and the degenerate approximation 
 for electrons (dashed-dotted line), for 
$ kT = 0.125 \; t$, and parameters explained in the text. 
The arrow marks the onset of negative curvature.}
\end{figure}

Let us consider a low concentration $(n)$ of carriers in the polaronic
configuration. The free energy at fixed temperature is proportional to the number
of carriers (polarons):
\begin{equation}                                                                
{\cal F}_{pol}(T,n) =   {\delta f}_{pol}(T)  \;   \;   \; n             
\end{equation}
where $ {\delta f}_{pol}(T)< 0  $ is the free energy reduction per carrier due
 to its polaronic bubble.
  The preference for polarons implies that the homogeneous
(paramagnetic) phase $ {\cal F}_{hom} $ is above $ {\cal F}_{pol}$ in the limit 
$ n \rightarrow 0$. 
 Consider now that the carrier density reaches the value $\tilde {n}$ at which
  all volume is
filled with polarons, each one with its confined carrier. It is obvious that, 
allowing carriers to delocalize (keeping core magnetization and temperature
constant),
 their kinetic energy diminishes. This
implies the existence of a homogeneous phase (with finite core magnetization
and total carrier polarization) with energy below the polaronic phase at finite
carrier concentration. A variational reasoning guarantees that the true
homogeneous solution of the  mean-field treatment is even below in free energy. 
In conclusion,
if the single polaron is stable, the free energy of the homogeneous phase  must
satisfy
\begin{eqnarray*}
{\cal F}_{pol}(T,n\rightarrow 0) = 
 {\delta f}_{pol}(T)  \;   \;   \; n 
& \; < \;&  {\cal F}_{hom}(T,n\rightarrow 0) \\
{\cal F}_{pol}(T,\tilde{n})=
  {\delta f}_{pol}(T)  \;   \;   \; \tilde{n}  
& \; > \; &  {\cal F}_{hom}(T,\tilde{n})
\end{eqnarray*}
Remenber that ${\cal F}_{pol}$ is linear in $n $, and both
 ${\cal F}_{pol}$ and ${\cal F}_{hom}$ start from the same point.
Then, the previous inequalities can only happen
if the curvature of $ {\cal F}_{pol}$ with concentration changes sign,
leading to a negative compressibility and PS. Therefore we have proven 
our original assertion: stable polarons imply thermodynamic instability towards
PS. Notice that this is a 
sufficient condition, but not necessary: PS can exists (see below) even if the 
ML polarons are not the stable configuration in the paramagnetic phase. A 
graphical
description of the previous argument comparing  polaronic and homogeneous
phase free energies for $kT=0.115  \; t$
is shown in fig.  \ref{figpiro1} (two upper curves). In addition, we also show
the free energy of the correct treatment,  including carriers entropy
(see below). 

There is, however, an important source of concern with the previous conclusion.
In the ML analysis, carriers in the homogeneous phase enter with energy but no 
entropy. This
would be correct for the usual case of degenerate fermions. But the temperature
is finite and the interesting region corresponds to very low carrier
concentration,
making the degenerate assumption questionable. For the parameters expected 
to apply in the
pyrochlores\cite{ML98}, the classical limit would be a more appropriate starting
 point.
 In fact, in the limit of zero
concentration at finite temperature, the chemical potential
 $(\mu = \partial_{n} {\cal F}_{hom})$
 corresponds to
classical particles, and diverges: $\mu( n \rightarrow 0) \sim \log(n)$.
 This implies that, in that limit, the homogeneous
paramagnetic phase (including carrier's entropy) is {\em always} more stable that the
ML polaronic ansatz.  Notice that our original statement remains true: 
 the existence of stable 
polarons would imply  
thermodynamic instability of the homogeneous phase. It only happens that the
carrier's entropy makes the paramagnetic homogeneous phase to be prefered 
over the polaronic one at low doping, making the very existence of polarons 
at finite doping uncertain.

Given the previous situation, we study the existence of PS in the pyrochlores 
 performing  the usual mean field calculation
{\em without} simplifying hypothesis for the carriers. The Helmholtz free 
energy $ {\cal F} $ contains the core  ${\cal F}_{\bar{s}} $, carrier
${\cal F}_{\bar{s}} $, and core-carrier contributions ${\cal F}_{s \bar{s}} $,
with expressions:
\begin{eqnarray}                                                                
{\cal F}_{s} &= &- \frac{J z}{2} s^2  - kT \; 
 \log \left[ \frac{{\rm sinh} (h)}{h}
 \right]  -h s    \nonumber \\
{\cal F}_{\bar{s}} &= &\sum_{\sigma=\uparrow \downarrow} \int d\epsilon \;
{\cal D}(\epsilon) \;
 \left\{ \epsilon - kT 
\log\left( e^{\frac{\epsilon-\mu_{\sigma}}{kT}} +1 \right)\right\}
  \nonumber \\
{\cal F}_{s \bar{s}} &= &- J' s \bar{s}           
\end{eqnarray}
where $z$ is the number of nearest neighbors, ${\cal D}(\epsilon)$ represents the 
 paramagnetic density of states (per  spin) and $h$ describes the
 effective field
characterizing the mean field distribution of core spins, assumed to be
classical vectors of unit length, therefore, $ s = {\rm ctnh}(h) -h^{-1}   $.

Minimization of ${\cal F} $ 
leads to the following equations:
\begin{eqnarray}                                                                
\mu_{\uparrow}  - J' s - \mu &=&  0   \label{piro1} \\
\mu_{\downarrow}  + J' s - \mu &=&  0 \label{piro2}   \\
-J z s + kT h - J' \bar{s} &=& 0       \label{piro3} 
\end{eqnarray}
with the occupation constraint:
\begin{equation}
n=\sum_{\sigma=\uparrow \downarrow} \int d\epsilon \;
{  {\cal D}(\epsilon) \over
  \exp({\epsilon-\mu_{\sigma}\over kT})  + 1 }
\end{equation}

\begin{figure} [!t]
\centering
\leavevmode
\epsfxsize=8cm
\epsfysize=8cm
\epsfbox {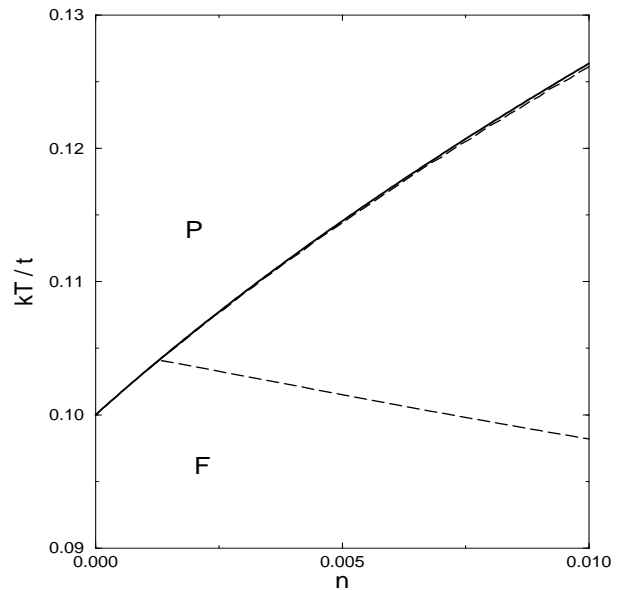}
\caption[]
{\label{figpiro2}
Phase diagrams  for the mean field theory of 
eqns. (\ref{piro1},\ref{piro2},\ref{piro3}) for the parameters explained in the
text.
The  solid line is the (scaled) critical
 temperature versus carrier concentration.
The region of negative compressibility is bound by the dashed line and the 
critical temperature.}
\end{figure}

We have solved the previous equations for parameters expected to apply in the
case of pyrochlores: 
$k T_0 = {Jz\over 3}= 0.1 \; t$, and $ J' = t $, where  $t$, 
 a measure  the
electronic energy scale, is taken to be the kinetic 
energy of an electron confined
to a  unit cell  volume. 
Searching for negative values of $\mu=\partial_n {\cal F}  $, we arrive at the
phase diagram plotted in fig.  \ref{figpiro2}   for small carrier concentration. We see that,  
beyond a  carrier concentration, a region of negative compressibility opens around the
ferro-para transition. Therefore, we confirm the existence of PS as suggested 
by the general arguments of section II. Notice that in  fig.  \ref{figpiro2} we show the area of
intrinsic instability: the standard Maxwell construction would increase this 
PS region even into the paramagnetic phase. 
We have plotted (fig.  \ref{figpiro1}) the free energy of the exact
solution (continuous line),  compared to the ML polaron ansatz (dashed, straight
line) and the
 solution ignoring the
electron entropy (dashed-dotted line) for   $kT=0.115  \; t$.  As explained
at the beginning of this section, the correct homogeneous phase
(including entropy) has the lowest free energy  even at low  concentration. 
In contrast, having 
ignored the electronic entropy would have led to the prediction that polarons are 
the prefered configuration (see two upper curves of fig.  \ref{figpiro1}).
Nevertheless, PS {\em does} exists, and the  onset of negative curvature (hardly
visible to the bare eye)
is marked with an arrow in that figure.

 For the parameters corresponding to fig.  \ref{figpiro2}, 
electrons are strongly 
 {\em non}
 degenerate, and the numerical results almost coincide  with the classical
 limit for the carriers, whose Ginsburg-Landau analytic treatment we now
 present for completeness.
  Assuming the electrons to be a classical ideal gas and performing an 
 expansion in the
  core $s$, and electron magnetization $ \bar{s}  $, one
 obtains the following expressions for the contributions to the free energy:
\begin{eqnarray}                                                                
{\cal F}_{s} &=& \left(\frac{3}{2} kT  - 
\frac{J z}{2}\right)\; s^2 + \frac{9}{20} kT\; s^4  
\label{fs}\\
{\cal F}_{\bar{s}}&=& -n k T +  n k T \; \log\left({n \over n_Q}\right)
+ {kT\over 2 n}\;{\bar{s}}^2  +  {kT\over 12 n^3}\;{\bar{s}}^4  \\
{\cal F}_{s \bar{s}}&=& - J' s \bar{s}           
\end{eqnarray}
where $ n_Q$ is a reference density marking the onset of the quantum regime.  

Minimization of the free energy leads to the following transition temperature:
\begin{equation}
kT_c = {kT_0\over 2} + \sqrt{\left({kT_0 \over 2}\right)^2+ {J'^2 n\over3}}
\end{equation}
 In the vicinity of $ T_c $, core and electron magnetization are related by   
\begin{equation}
\bar{s} = s {J' n \over kT } \; \left[ 1- { J'^2  s^2\over 3 (kT)^2}\right]
\end{equation}
 and the  scaled  inverse compressibility at $T\rightarrow T_c^{-}$ is given by:
\begin{equation}
\tilde{\kappa}^{-1} \equiv
\frac{\partial^2 {\cal F}_{MF}}{\partial n^2}
 = \tilde{\kappa}_0^{-1} -
{ 15 \; kT_c \; J'^4  \over 54 (kT_c)^4  + 10 J'^4 n   }
\end{equation}
where $\tilde{\kappa}_0^{-1}=kT_c / n$, is the inverse compressibility of the paramagnetic phase.
  $\tilde{\kappa}$ becomes negative beyond  {$ n_{PS} = { 54 T_c ^4 / (5J'^4 )} =
  1.266 \times 10^{-3}$}, for the parameters used here,
very close to the exact results of fig.  \ref{figpiro2}.
 
Aa a final remark, it is important to
realize that the precise configuration of the 
coexisting
phases in the real material will be  complicated by effects beyond the present 
treatment (disorder, domain boundary contributions ...). Nevertheless, the associated spatial
inhomogeneity will certainly affect the physics of these compounds in an
important way. This strongly suggests that, even if the original polaron
argument is doubtful, the associated physics explored by ML might well apply
to the pyrochlores.

\section{Double exchange systems.} 
The double exchange (DE) model describes systems with local spins
and itinerant electrons, in the limit where the Hund coupling
between the electrons and the spins is much larger than
other scales\cite{Z51}. The electrons are always polarized
in the direction of the local spins, and hopping $(t)$ to neighboring
ions is modulated by the relative orientation of these spins, leading to
the following Hamiltonian
\begin{equation}
{\cal H} = -t\sum_{\langle ij\rangle,\sigma}
\left[ z ^{\vphantom{\dagger}}_{i\sigma}
{\bar z} ^{\vphantom{\dagger}}_{j\sigma}\,
c ^\dagger_i c ^{\vphantom{\dagger}}_j + {\rm H.c.}\right]\ .
\label{hamil}
\end{equation}
where $z ^{\vphantom{\dagger}}_{i\sigma}$ is the spinor
describing the orientation of the core at site $i$:
$z ^{\vphantom{\dagger}}_{i \uparrow}=\cos( \frac{1}{2}\theta_i)$,
$z ^{\vphantom{\dagger}}_{i \downarrow}=
\sin( \frac{1}{2}\theta_i)\,\exp(-i\phi_i)$.

 The model, put
forward by Zener\cite{Z51}, has received great attention in recent times as
 the 
building block needed to explain the physics of $Mn$ perovskites. We will 
restrict ourselves to its simplest one-band version, leaving aside
further features such as orbital degeneracy, antiferromagnetic couplings 
or Jahn-Teller distortion, probably required for a more realistic
description of manganites.  

PS  in double exchange models has been extensively 
studied\cite{N96,RHD97,AG98,Yetal98,Detal98,AGG98,MYD99}.  It was originally
observed in numerical studies of the DE model with  AF couplings
between core spins,  raising doubts  about its existence 
in the bare version\cite{RHD97,Yetal98,Detal98}. However, we have recently
 shown that PS tendency 
is an intrinsic feature of  the simplest DE  model without additional
terms\cite{AGG98}.
 For completeness,
we include here an analysis of the compressibility using the
Ginzburg-Landau expansion outlined in the preceding sections.
The approach is a direct extension of that used in\cite{G60}.
We leave out the study of PS at zero temperature
due to the competition between direct antiferromagnetic 
interactions and the double exchange mechanism\cite{AG98},
which also can be cast in terms of a Ginzburg-Landau expansion. 

 The model is determined by the polarization 
 of the localized spins,
$s$, the density of states of the itinerant electrons in the
absence of spin disorder, ${\cal D}_0( \epsilon )$, and the temperature $T$.
The free energy of
the spins comes from the entropy due to their thermal
fluctuations only, and coincides with eq.(\ref{fs}), except
for the absence of the direct exchange term.
 We assume that the electron gas
is degenerate, an excellent approximation for temperatures in the range
of $T_c$, and replace its free energy by the ground state
energy in a background of fluctuating spins. The bandwidth of
the electrons is reduced by a factor $f = \langle \cos ( \vartheta{ij} / 2 )
\rangle$, where $\vartheta{ij}$ is the angle between neighboring spins.
Thus, the electronic energy can be written as:
\begin{equation}
E = f K_0 = f \int_{- W_0}^{\mu_0} \epsilon  \; 
{\cal D}_0 ( \epsilon ) d \epsilon
\end{equation}
where $\mu_0$ denotes the chemical potential in the
absence of spin disorder,
and $W_0$ is the lower band edge
(note that $- W_0 \le \mu_0 \le W_0$). We can now expand $f$ in terms of
the magnetization $s$, to obtain:
\begin{equation}
E = \frac{2}{3} K_0 + \frac{2}{5} K_0 s^2 -
\frac{6}{175} K_0 s^4
\label{kineticmf}
\end{equation}
which completes the required Ginzburg-Landau expansion.
The Curie temperature of the model is given by $T_c =
- \frac{4}{15} K_0$ ($K_0$ is negative). The magnetization is:
\begin{equation}                                                                
s^2 = \left\{ \begin{array}{lr}                                                 
0 &T \geq T_c \\                                                                
-\frac{\frac{3}{2} kT + \frac{2}{5} K_0}{\frac{9}{10} kT - \frac{12}{175}          
K_0} &T < T_c \end{array} \right.
\label{spindemf}                                               
\end{equation}  
which leads to:
\begin{equation}                                                                
{\cal F} = \left\{ \begin{array}{lr} \frac{2}{3} K_0 &T\geq T_c \\              
\frac{2}{3} K_0 - T f \left( \frac{K_0}{T} \right)                              
& T < T_c \end{array} \right.                                                   
\end{equation}                                                                  
where                                                                          
$f ( u ) = \frac{7}{6} \frac{\left( u + \frac{15}{4} \right)^2}{                
\frac{105}{8} - u}$.

Finally, the scaled inverse compressibility is:
\begin{equation}                                                                
\tilde{\kappa}^{-1} = \left\{ \begin{array}{lr} \frac{2}{3 {\cal D}_0 ( \mu_0 )} &T \rightarrow     
T_c + 0 \\ \frac{2}{3 {\cal D}_0 ( \mu_0 )} - \frac{14 \mu_0^2}{27 | K_0 |}            
&T \rightarrow T_c - 0 \end{array} \right.                                      
\end{equation}                                                                  
The compressibility below $T_c$ is negative when:                               
\begin{equation}                                                                
\frac{| K_0 |}{{\cal D}_0 ( \mu_0 )} \leq \frac{7}{9} \mu_0^2                          
\end{equation}                                                                  
This result is independent of the initial bandwidth, $2 W_0$.                   
In general, there is always PS at low                             
fillings, because $\mu_0 \rightarrow - W_0$ while                               
$\frac{K_0}{{\cal D}_0 ( \mu_0 )} \rightarrow 0$. At half filling,                     
$\mu_0 \rightarrow 0$, and there is no PS.  
                      
If we use:                                                                      
\begin{equation}                                                                
{\cal D}_0 ( \epsilon ) = \frac{2}{\pi} \frac{\sqrt{W_0^2 - \epsilon^2 }}{W_0^2}
\label{bethe}       
\end{equation}                                                                  
we find that there is PS if:                                      
\begin{equation}                                                                
| \mu_0 | \geq W_0 \sqrt{\frac{3}{10}}                                          
\end{equation} 
which corresponds to a number of electrons per site $\sim 0.2$.

The previous  expansion in powers of the order parameter is not 
necessary, and the complete mean-field equations can be written as follows
\cite{AGG98} 
\begin{eqnarray}
x - {1\over 2}&=&{1\over 2}\int_{-1}^1\!\!\!\! d \gamma\, {\cal D}_0( \gamma)\,\tanh
\left({\alpha +{f}  \gamma\over 2\Theta}\right)\label{dg1}\\
2\Theta\,{ {f}Q\over s}&=&
{1\over 2}\int_{-1}^1\!\!\!\! d \gamma\, \gamma\, {\cal D}_0( \gamma)\,
\tanh\left({\alpha +{f}\gamma \over 2\Theta}\right)
\label{dg2}
\end{eqnarray}
where $s= \,{\rm ctnh\,}(Q)-Q^{-1}$ is the magnetization, $x$ describes the
doping level (electron/hole),
and ${f}=\sqrt{ \frac{1}{2}(1+s^2)}=\langle \cos^{2}(\vartheta{ij}) \rangle
^{1/2}$.
The energy, temperature and chemical potential are scaled in units of the
half-bandwidth in the absence of spin disorder  
by $\gamma \equiv \epsilon/W_0 $, $\alpha\equiv\mu/W_0$, 
and $\Theta\equiv kT/W_0$.

\begin{figure} [!t]
\centering
\leavevmode
\epsfxsize=8cm
\epsfysize=8cm
\epsfbox {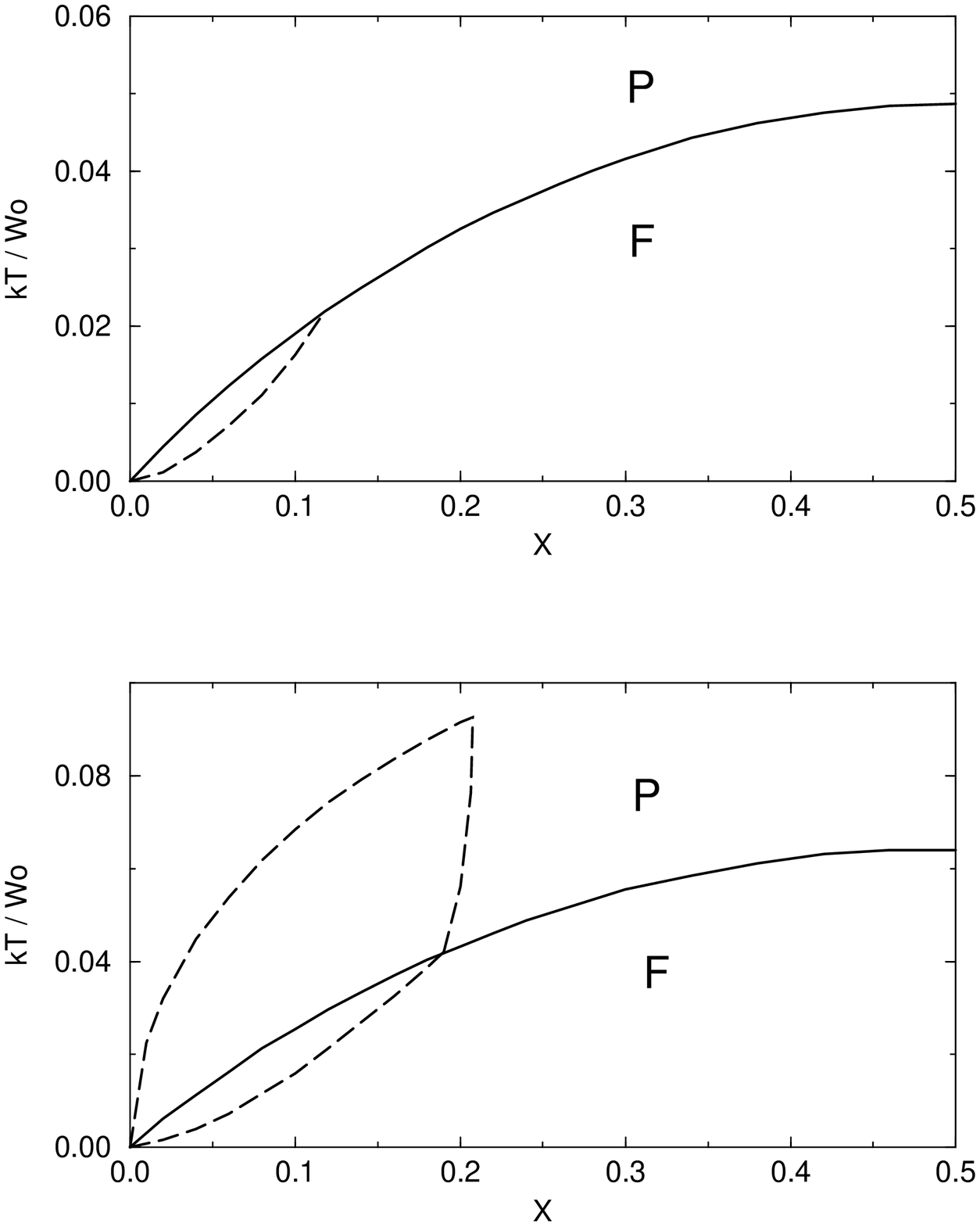}
\caption[]
{\label{defig1}
Phase diagrams (upper panel) for the mean field theory of 
eqns. (\ref{dg1},\ref{dg2}),
and (lower panel) for that of eqns. (\ref{hmf},\ref{sbmf}), 
both computed using an
elliptic density of states.  The  solid line is the (scaled) critical
 temperature versus hole concentration.
The  dashed line marks the boundary of the
region of negative compressibility.} 
\end{figure}

If ${\cal D}_0$ is  taken to be the density of states of the original crystal
 lattice,  neglect of  the Berry phase\cite{berry} collected by the 
 hopping electron is implicit. More appropriately, we can assume that
spin disorder cancellations favor retraced paths\cite{br70}, leading to 
a Bethe lattice density of
 states, with eq.(\ref{bethe}) as the infinite coordination limit. 
 In fig.  \ref{defig1} (upper panel) we show the
 phase diagram (scaled temperature versus hole concentration $x$) for the density of eq.(\ref{bethe}). Notice that
 the PS region corresponds to the intrinsic instability (negative
compressibility). 
 Global stability (Maxwell construction) would  increase this region.

 The simple DE model offers a clear example of the general mechanism for PS
 in the presence of an ordering process. In fact, the mean field treatment
 is intrinsically unstable versus PS in the limit of low carrier density. To
 see this, notice that the energy scale of the magnetic order is the
 electronic energy, proportional to the carrier concentration in the dilute limit.
 Doubling, for instance, the carrier concentration close to the Curie temperature
 produces a  finite increase of the bandwidth. The chemical potential
instead
 remains tied to the band edge and follows its fate: it decreases with 
 increasing electron density (thermodynamic instability). In our electron-hole
  symmetric model, the same would apply to the dilute hole limit.

  Although we do not expect the mean field approximation to be a serious
  drawback, certain limitations are clear. For instance, the inability 
  to distinguish between short-range correlations and long-range order 
  confines the negative compressibility region to the ferromagnetic phase,  
  for only there does the bandwidth change with magnetic order. In what 
  follows, we will show that PS is a robust feature that survives several
  improvements of this basic mean field approach.

 \subsection{Schwinger bosons.}

An improved mean field theory can be implemented using the
Schwinger boson \cite{aa88,AGG98} method.  A nondynamical field $\lambda_i$
enforces the constraint ${\bar z}^{\vphantom{\dagger}}_{i\sigma}
z^{\vphantom{\dagger}}_{i\sigma}~=~1$ at every site.  The core spins are
quantized according to $[z ^{\vphantom{\dagger}}_{i\sigma},
{\bar z} ^{\vphantom{\dagger}}_{j\sigma'}]=
\delta_{ij}\delta_{\sigma\sigma'}/2S$.
The mean field Hamiltonian is obtained through a Hartree decoupling of
the bosonic $z ^{\vphantom{\dagger}}_{i\sigma}
{\bar z} ^{\vphantom{\dagger}}_{j\sigma}$ and
fermionic $c ^\dagger_i c ^{\vphantom{\dagger}}_j$
hopping terms:
\begin{eqnarray}
{\cal H}_{\scriptscriptstyle{\rm MF}}&=&N(ztfK - \lambda) -\mu\sum_i
c ^\dagger_i c ^{\vphantom{\dagger}}_i
+\lambda\sum_{i,\sigma}  {\bar z} ^{\vphantom{\dagger}}_{i\sigma}
z ^{\vphantom{\dagger}}_{i\sigma}
\label{hmf}\\
&&-tf\sum_ {\langle ij\rangle} (c ^\dagger_i c ^{\vphantom{\dagger}}_j
+ c ^\dagger_j c ^{\vphantom{\dagger}}_i) -tK
\sum_{ {\langle ij\rangle},\sigma}
({\bar z} ^{\vphantom{\dagger}}_{i\sigma} z ^{\vphantom{\dagger}}_{j\sigma}
+ {\bar z} ^{\vphantom{\dagger}}_{j\sigma}
z ^{\vphantom{\dagger}}_{i\sigma})\ ,\nonumber
\end{eqnarray}
where $K=\langle c ^\dagger_i c ^{\vphantom{\dagger}}_j\rangle$
and $f=\langle z ^{\vphantom{\dagger}}_{i\sigma}
{\bar z} ^{\vphantom{\dagger}}_{j\sigma}\rangle$
Such a model was introduced by Sarker \cite{s96}, who identified a Curie
transition and found that the $e_g$\ fermion band becomes incoherent
above $T_ {\rm c}$. 
Accounting for the possibility of condensation of Schwinger bosons, we write
$\Psi_{{\fam\mibfam\eightmib k}\sigma}\equiv\langle
z^{\vphantom{\dagger}}_{ {\fam\mibfam\eightmib k}\sigma}\rangle$.  Assuming
condensation only at $ {\fam\mibfam\tenmib k}=0$, we define
$\rho\equiv |\Psi_{ {\fam\mibfam\eightmib k}=0,\sigma}|^2$.
The mean field equations are then
\begin{eqnarray}
1+{1\over 2S}&=&\rho+{1\over 2S}\int_{-1}^1\!\!\!\!
d \gamma\,{\cal D}( \gamma)\,
{\rm ctnh}\!\left({\Lambda-K\gamma\over 4S\Theta}\right)\nonumber\\
f&=&\rho+{1\over 2S}\int_{-1}^1\!\!\!\!
d \gamma\,\gamma\,{\cal D}( \gamma)\,
{\rm ctnh}\!\left({\Lambda-K\gamma\over 4S\Theta}\right)\nonumber\\
x - {1 \over 2}&=&{1\over 2}\int_{-1}^1\!\!\!\! d \gamma\,{\cal D}( \gamma)\,
\tanh\!\left({\alpha + f \gamma \over 2\Theta}\right)\nonumber\\
K&=&{1\over 2}\int_{-1}^1\!\!\!\! d \gamma\, \gamma\,{\cal D}( \gamma)\,
\tanh\!\left({\alpha +f \gamma\over 2\Theta}\right)\ ,
\label{sbmf}
\end{eqnarray}
where  $ \gamma = \epsilon / W_0 $, 
$\Lambda=\lambda/W_0$, $\Theta= k_{ \scriptscriptstyle {\rm B}}T/W_0$,
and $\alpha=\mu/W_0$,  $W_0$ being the half-bandwidth in the absence of spin
disorder. 

Two aspects distinguish this treatment from that of the previous section.
First, core spins are quantum objects with intrinsic dynamics: a minute  
correction close to the Curie temperature. Second and more important,  
long-range magnetic order (condensation of Schwinger bosons) is not a
requisite for 
short range correlations: the latter exist and change with temperature 
even in the paramagnetic phase.

We have solved the mean field equations for a semi-elliptic density of
states  and $S=3/2$ \cite{AGG98}.  Several features of the solution are shown in
fig.  \ref{defig1} (lower panel). While the critical temperature is hardly 
affected, the   region of negative
compressibility enters now into the paramagnetic phase,
an expected physical feature not present in the previous approximation.
In this approach, the fermion bandwidth collapses to zero at high temperature.
Although this is likely an artifact of the approximations, it takes place well
above the Curie temperatures to be a source of concern.

 \subsection{Critical spin fluctuations}

  The rapid change of spin correlations  ties the 
PS region to the magnetic ordering transition. One can worry about the
importance of critical spin fluctuations, neglected in the molecular field 
approximation. 
 Assuming a nearest-neighbors, tight-binding  lattice
with
hopping amplitude $t_{ij}=-t \cos(\theta_{ij}/2)$, and expanding the
half-angle between
spins in the usual manner: $\cos(\theta/2) \simeq a_0 + a_1 cos(\theta) $,
the standard  virtual
crystal decoupling  of the electron-spin system allows   integration 
of charge degrees of freedom . This produces an effective Heisenberg model for
the  core spins (assumed classical for simplicity) and the total free
energy (charge and core spins) is  given by
\begin{equation}
{\cal F} = K_{\infty}(x) + {\cal F}_{H}(J,T)
\end{equation}
where $ K_{\infty} $ is the electron energy (degenerate limit assumed) 
of the fully disordered limit
$ t_{ij}=a_0 t $.
${\cal F}_{H} $ describes
the free energy of the classical Heisenberg model, with coupling
constant $J=2 a_1 |\langle c^{\dagger}_i c_j\rangle|t $, where tilded quantities are
electron properties scaled to unit half-bandwidth.  A lengthy but
simple calculation permits de following connection between the stability
criterium and  thermal properties of the Heisenberg model (cubic lattice):
\begin{equation}
\frac{\partial^2 {\cal F}}{\partial n^2} = 
\frac{1}{{\cal D}(\mu)}
\left[ \: 1 - g(x) \; f_H(T) \: \right]
\end{equation}
with
\begin{equation}
g(x) = \frac{2 a_1 \tilde{\mu}^2 \tilde{\cal D}(\tilde{\mu})}
{ 6 | \langle c^{\dagger}_i c_j\rangle |}
\end{equation}
and
\begin{equation}
 f_H(T) = \frac{T \; C(T)}
 {J \;\left[ \: a_0 + a_1 \langle \cos(\theta_{ij})\rangle \:\right]} 
\end{equation} 
 $\mu$ is the chemical potential and $x$ the carrier density.
$C(T)$  represents the specific heat of the Heisenberg model.  
The expected increase  of $C(T)$ near the 
critical temperature $T_c$ links the tendency towards instability with
magnetic ordering. In any case, for a fixed value of $T/T_c$, the
system
is bound to exhibit PS, $(\partial \mu / \partial x) < 0$, 
owing to the diverging behavior of $g(x\rightarrow 0) $. Extracting
 the temperature dependence of $f_H(T)$ from published
Monte Carlo data\cite{lau} for the cubic lattice,
 we estimate PS accompanying the ordering
transition
up to $x\simeq 0.11$.
The resulting phase diagram and intrinsic instability region are shown in
fig.   \ref{defig2} (upper panel). Notice that no approximate
treatment is assumed
for the criticality of the effective magnetic Hamiltonian, the only
(major) surviving approximation being the  virtual crystal (mean field) 
decoupling of mobile carriers and core spins.

\subsection{Infinite dimension} 
 
The mean field decoupling of charge and core spins fluctuations remains in 
all the previous analysis. This manifests itself in a mere rescaling of the
electronic density of states with magnetic order. As the PS instability can be
 seen as a manifestation 
of correlation between charge and spin at a macroscopic level (the only possibility
left in a mean field approach) concern might arise about its fate when 
 charge-spin coupling is properly considered. Although in the
general case a numerical approach would be necessary to answer this question, there
is a limit in which the  problem is solvable but the charge-spin coupling remains
non trivial: infinite coordination (dimension). 
  The relevance of this  limit for the understanding of correlated systems in 
  3-D  has been recognized in recent years \cite{MV89,GKKR96}. 

 In the DE case, this problem has been 
 considered by Furukawa \cite{f95}, and indeed PS has been
  observed \cite{Yetal98,Detal98}.
 In those works,  the DE model included a large though finite Hund coupling,
 leading to AF correlations between core spins that certainly
 increase the tendency towards PS. This fact, together with accummulated
 experience from numerical studies of doped antiferromagnets, has led to the
 extended belief that antiferromagnetism is required for PS to exist. We now
 show that, in spite of recent suggestions on the contrary,  
 the simple DE 
 model without AF additions  indeed exhibits PS close to the ordering
 temperature, according to the general explained in this paper.

\begin{figure} [!t]
\centering
\leavevmode
\epsfxsize=8cm
\epsfysize=8cm
\epsfbox {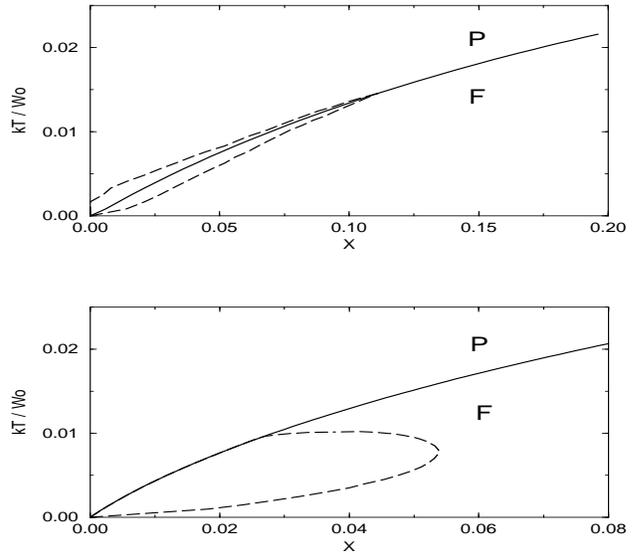}
\caption[]
{\label{defig2}
Phase diagrams (upper panel) for the cubic lattice in the virtual crystal approach, including
MonteCarlo data for the Heisenbeg model (subsection B) and (lower panel) for the Bethe
lattice of infinite coordination.  The 
 solid line is the (scaled) critical temperature
versus
hole concentration.
The dashed line marks the boundary of the
region of negative compressibility.} 
\end{figure}

 The average density of states for electrons moving in an infinitely coordinated
 Bethe lattice according to the DE Hamiltonian is given by: 
 \begin{equation}
 \langle {\cal D}(\epsilon) \rangle_{\vec{\Omega}} =  -(1/ \pi) \;  \Im \;
 \langle g(\epsilon,\vec{\Omega}) \rangle_{\vec{\Omega}} 
 \end{equation}
 where the local Green's function satisfies:
 \begin{equation}
 g(\epsilon,\vec{\Omega}) = 
 \left[ \; \epsilon- \Sigma(\epsilon,\vec{\Omega})\;\right]^{-1}
 \end{equation}
  and the self-energy is determined   by:
 \begin{equation}
\Sigma(\epsilon,\Omega) = ( W_0/2) ^ 2 \;
 \langle  \: \frac{1 + \vec{\Omega} \cdot \vec{\Omega}^{\prime}}{2} \;
 \left[\epsilon - \Sigma(\epsilon,\vec{\Omega}^{\prime})\right]^ {-1} \:
  \rangle_{\vec{\Omega}^{\prime}} 
  \label{inf1}
 \end{equation} 
 $\vec{\Omega}$ 
 is the unit vector describing the local core  (classical) spin,
 $W_0$ represents the half-bandwidth for the fully aligned case, and angular 
 averages are
 taken with the  probability distribution 
 ${\cal P}(\vec{\Omega})$. Notice that, though the probability distribution for 
 spins in different sites factorizes (confining PS the the ferromagnetic
 phase), 
  charge and spin fluctuations remain coupled at the same site, and the  density
   of states 
  is not  merely rescaled by magnetic order (eq. \ref{inf1}).

  The complete thermodynamic problem is solved  by finding the probability distribution
  that minimizes the total free energy.
  \begin{equation}
  {\cal F} = {\cal F}_{el} - T {\cal S}
  \end{equation}
  where $ {\cal F}_{el} $ is the free energy of fermions for the  average density 
  of states
  and   ${\cal S} $ is the core spin entropy, both depending on 
   ${\cal P}(\vec{\Omega})$.

\begin{figure} [!t]
\centering
\leavevmode
\epsfxsize=8cm
\epsfysize=6cm
\epsfbox {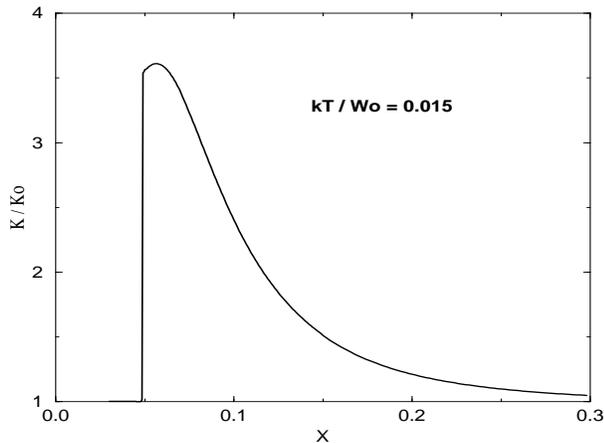}
\caption[]
{\label{defig3}
Compressibility of the DE model for the (infinitely coordinated)
 Bethe lattice, in units of the
compressibility of a free fermion system with identical density
 of states, versus hole
concentration at constant temperature.} 
\end{figure}
  
  For computational convenience, we have solved the problem with two 
  additional
  features: degenerate limit for carriers and a  core spin distribution
  parametrized by an effective magnetic field (detailed analysis of the 
  exact
  solution for a few temperatures shows the previous simplifications to 
  introduce only minute corrections).  
  The generic results of 
  previous treatments are reproduced here. This is shown  
  in fig.   \ref{defig2} (lower panel), where the phase
  diagram  presents the expected region of intrinsic instability 
  (negative compressibility)  at low carrier density. 
  The inclusion of a finite Hund coupling will certainly increase the PS region,
  as observed in previous works \cite{Yetal98,Detal98}. 
   All these  studies support the view that  the general scheme for PS 
   presented in this
  paper applies to the simple DE model with features which are robust and 
  not an  artifact of approximations.

 As mentioned before, PS can be thought of as a long range instability 
 caused by the coupling between carriers and core spins. Such mechanism is 
 operative even in the region where the system does not phase separate,  and 
 the  carriers compressibility can be much enhanced close to the Curie 
 temperature. 
 This statement does not contradict the fact that the density of states 
 at the Fermi level is featureless and shows no major change around the Curie
 temperature \cite{cb98}. Remember that, if the bandwidth were not affected by the spin
 order, the compressibility would be merely proportional to the  density of states at 
 the Fermi level, that means featureless. It is the  bandwidth change with
 magnetic order what produces this enhancement 
 in the region close to the onset of magnetic order. This is shown in Fig
 \ref{defig3},  
 where the compressibility of the infinitely coordinated Bethe lattice  is 
 measured in units of the compressibility of a {\em free} fermion system (that
 is,
 no carrier-spin coupling) with the same density of states. This behavior is
 not a peculiarity of the infinitely dimension limit, and remains the
 the same in all  previous approaches. The possible connection between 
 this enhancement and the unusual properties of the manganites remains an open
 and intriguing question \cite{MYD99}.

\section{Influence of a magnetic field}
So far, we have considered the existence of phase separation 
near a ferromagnetic (or antiferromagnetic) - paramagnetic transition,
at zero magnetic field. When a field is applied, the features
associated to the transition are smoothed, and, at sufficiently
large fields, the magnetic moments are aligned, and the magnetization
has a weak temperature dependence.

The increase in the compressibility which leads to phase separation is
asociated with the strong coupling between magnetic and charge
fluctuations, as measured by the dependence of the critical
temperature on electronic concentration. The magnetic field
suppresses magnetic fluctuations leading to a smaller
increase in the compressibility near $T_c$. 
Phase separation should dissappear at sufficiently large
fields, when the temperature no longer induces significant
magnetic fluctuations.

\begin{figure} [!t]
\centering
\leavevmode
\epsfxsize=8cm
\epsfysize=6cm
\epsfbox {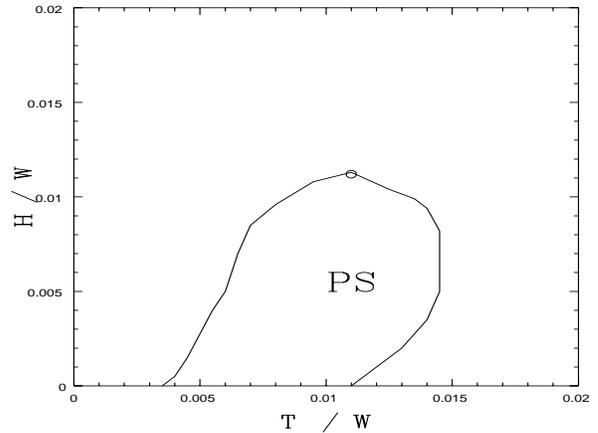}
\caption[]
{\label{field}
Region of phase separation,
as function of magnetic field, in the double exchange model, for
an electron density, $x=0.10$.
Temperature and magnetic field are normalized to the
electron bandwidth.}
\end{figure}

A generic case which shows the dependence of the region where
phase separation occurs on magnetic field,
at fixed nominal electronic concentration, is shown
in fig. (\ref{field}). The calculations have been done
for the double exchange model of the previous sections,
using the mean field equations (\ref{kineticmf},
\ref{spindemf}),
plus an applied field. The region of negative compressibility
is shown. As in previous examples, a Maxwell construction
gives a somewhat larger region.

In a small applied field, phase separation takes place above the
Curie temperature. This is due to the rounding of the discontinuity
at $T_c$ of the compressibility induced by the field.
The region of phase separation lies between a high magnetization and
a low magnetization phase, in close analogy with the
phase diagram of an ordinary liquid-vapor transition.
At high fields, phase separation is completely suppressed.
At the highest possible field we find a critical point. 
  
\section{Phase separation and domain formation}
The analysis in the previous sections suggests that PS should be
a frequent feature of magnetic transitions. PS on
a macroscopic scale, however, cannot occur, as it requires an infinite
amount of electrostatic energy.
We can extend the mean field analysis used in the previous section
to study the role of the Coulomb interaction. If we first neglect it,
the charge susceptibility obeys:
\begin{equation}
\lim_{q \rightarrow 0}{\raise.35ex\hbox{$\chi$}}_0(q)
=-{\cal V}^{-1}\frac{\partial n}{\partial \mu} = -
{\cal V}^{-1} n^2 \kappa
\end{equation}
where ${\cal V}$ is the volume of the unit cell.
A charge fluctuation $\rho ( \vec{q} )$ induces an electrostatic potential,
$( 4 \pi e^2 / q^2 )\rho ( \vec{q} )$, which induces more
charge polarization. The RPA selfconsistent equation for the
charge polarizability becomes:
\begin{equation}
\lim_{q \rightarrow 0}{\raise.35ex\hbox{$\chi$}}_(q) =
\frac{{\raise.35ex\hbox{$\chi$}}_0(q)}{1 - \frac{4 \pi e^2}{q^2}
{\raise.35ex\hbox{$\chi$}}_0(q)}
\rightarrow \frac{-{\cal V}^{-1} n^2 \kappa}{1 + \frac{4 \pi e^2}{q^2}
{\cal V}^{-1} n^2 \kappa}
\end{equation}             
when, in the absence of Coulomb repulsion the system  exhibits the
instability,
$\kappa < 0$, the denominator has a pole at:
\begin{equation}
q_*=\sqrt{-{4\pi e^2\over{\cal V}}\,{ n^2 \kappa}}\ .
\end{equation}
and charge fluctuations of shorter wavelengths remain unstable, whereas 
the long range nature of the Coulomb term prevents the formation of larger 
charge inhomogeneities.
Thus, we expect the formation of domains at length scales
comparable to $q_*^{-1}$. 
As mentioned previously, this analysis does not take into account neither 
the cost in magnetic energy associated with the formation of
domain walls nor the effect of impurities.
In addition to macroscopic charge neutrality, these are likely to be among 
the major factors affecting the spatial coexistence pattern in real materials.

\section{Conclusions}
We have discussed a general framework which shows that PS
is likely to occur near magnetic phase transitions. The dependence
of the critical temperature, or critical couplings, on electronic
density leads to a reduction to the compressibility in the
ordered phase. This reduction tends to be of the same order as the
value of the compressibility in the disordered phase. The existence
of PS typically depends on numerical constants of order unity,
related to the electronic structure. The coupling constants do
not require a special fine tuning for PS to take
place. The analysis presented here is probably relevant to
understand a variety of experimental findings in the
manganites, the pyrochlores and doped antiferomagnets.

We have studied PS mostly within a conventional
mean field framework. Corrections due to critical fluctuations
will usually enhance the effects discussed here. The compressibility
is associated with the second derivative of the free energy 
with respect to electronic density, in a similar way to
the specific heat. Critical fluctuations tend to
increase the divergences of the derivatives of the free energy
at the transition, so that the estimates reported here 
probably underestimate the tendency towards PS.
The analysis also shows that, even in the absence of phase
separation, an enhanced coupling between charge and magnetic
fluctuations is expected, due to the reduction in the
charge compressibility.

Finally, we have discussed the way in which electrostatic effects
frustrate the formation of macroscopic domains, and we give
a scheme to calculate the scales at which domain formation is 
expected.
\section{Acknowledgements.}
We are thankful to D. Khomskii for many illuminating discussions.
Financial support through grants PB96-0875 (MEC, Spain)
and (07N/0045/98) (C. Madrid) and the Comission for Scientific Exchange
between Spain and the USA are gratefully acknowledged.


\begin{references}
\bibitem{K98}
We are thankful to D. Khomskii for pointing this out to us.
\end{references}
\end{document}